\documentclass[conference]{IEEEtran}
\IEEEoverridecommandlockouts

\usepackage{amsmath,amssymb,amsfonts}
\usepackage{algorithmic}
\usepackage{graphicx}
\usepackage{textcomp}
\usepackage{xcolor}
\usepackage{subfigure}

\usepackage[style=ieee,backend=biber,bibencoding=utf8]{biblatex}
\renewcommand*{\bibfont}{\footnotesize}

\usepackage{numbers}

\newcommand{\focus}{\textit{in\textbar FOCUS}}

\newcommand{\explorviz}{\textit{ExplorViz}}

\newcommand{\adesso}{\textit{adesso SE}}

\def\BibTeX{{\rm B\kern-.05em{\sc i\kern-.025em b}\kern-.08em
    T\kern-.1667em\lower.7ex\hbox{E}\kern-.125emX}}

\graphicspath{{img/}}

\makeatletter
\newcommand{\linebreakand}{%
\end{@IEEEauthorhalign}
\hfill\mbox{}\par
\mbox{}\hfill\begin{@IEEEauthorhalign}
}
\makeatother

\begin{document}

\title{Microservice Decomposition via Static and Dynamic Analysis of the Monolith}

\author{\IEEEauthorblockN{Alexander Krause}
\IEEEauthorblockA{\textit{Software Engineering Group} \\
\textit{Kiel University}\\
Kiel, Germany \\
akr@informatik.uni-kiel.de}
\and
\IEEEauthorblockN{Christian Zirkelbach}
\IEEEauthorblockA{\textit{Software Engineering Group} \\
\textit{Kiel University}\\
Kiel, Germany \\
czi@informatik.uni-kiel.de}
\and
\IEEEauthorblockN{Wilhelm Hasselbring}
\IEEEauthorblockA{\textit{Software Engineering Group} \\
\textit{Kiel University}\\
Kiel, Germany \\
wha@informatik.uni-kiel.de}
\linebreakand
\IEEEauthorblockN{Stephan Lenga}
\IEEEauthorblockA{\textit{Software Engineering Group} \\
\textit{Kiel University}\\
Kiel, Germany \\
stephan\_lenga@web.de}
\and
\IEEEauthorblockN{Dan Kröger}
\IEEEauthorblockA{\textit{Solution Center Lottery Applications} \\
\textit{adesso SE}\\
Hamburg, Germany \\
dan.kroeger@adesso.de}
}

 \maketitle

\begin{abstract}
Migrating monolithic software systems into microservices requires the application of decomposition techniques to find and select appropriate service boundaries.
These techniques are often based on domain knowledge, static code analysis, and non-functional requirements such as maintainability.

In this paper, we present our experience with an approach that extends static analysis with dynamic analysis of a legacy software system's runtime behavior, including the live trace visualization to support the decomposition into microservices. 
Overall, our approach combines established analysis techniques for microservice decomposition, such as the bounded context pattern of domain-driven design, and enriches the collected information via dynamic software visualization to identify appropriate microservice boundaries.

In collaboration with the German IT service provider adesso SE, we applied our approach to their real-word, legacy lottery application \focus~to identify good microservice decompositions for this layered monolithic Enterprise Java system.
\end{abstract}


\section{Introduction}
Improved maintainability, short time to market, and high scalability are some benefits of the microservice architectural style~\cite{NewmanBuildingMicroservices}.
These advantages act as drivers for companies to modernize monolithic software systems towards microservices.
While there are barriers for a microservice adoption~\cite{KnocheDriversMicroservices,Carrasco:2018}, they are an often desired architecture for migrating monolithic software systems to cloud-native environments~\cite{MigrationToMicroservicesSurvey:2018}.

These migration processes are rarely started as greenfield projects due to cost and time constraints~\cite{NewmanBuildingMicroservices}.
Instead, monolithic software systems are incrementally decomposed into microservices.
The decomposition, however, is a challenging task and requires many iterations to find suitable service boundaries~\cite{FowlerMonolithInMicroservices,NewmanBuildingMicroservices}.
Industry best practices~\cite{MicrosoftDomainAnalysis,MicrosoftMicroserviceBoundaries,InnoqDecompositionStrategies} and studies from academia~\cite{MigrationToMicroservicesSurvey:2018} introduce strategies to support this task, but often share the same techniques, e.g., bounded contexts (BC) of the domain-driven design (DDD)~\cite{evans2004ddd}, static code analysis, and refactoring based on non-functional requirements.

In this paper, we present our approach for microservice decomposition which additionally includes a dynamic analysis and visualization of a software system's runtime behavior to find potential microservice boundaries.
Furthermore, we report on its application to a real-world, legacy monolithic online lottery software called \focus.\footnote{\url{https://www.adesso.de/en/adesso-branch-solutions/lotteriegesellschaften/leistungen/loesungen/}}
\focus\ is a sales and management software solution for lottery providers that is developed by \adesso, one of the largest IT service providers in Germany.
It is provided as a software as a service (SaaS) solution which is currently used by several state lotteries.

Our approach for microservice decomposition starts with established migration processes including a domain analysis of \focus~to familiarize with its ubiquitous language and business domain~\cite{FowlerMonolithInMicroservices,MicrosoftDomainAnalysis}.
Based on the outcome, we selected bounded contexts, i.e., the foundation for decomposing the layered monolithic Enterprise Java system into microservices when using DDD~\cite{NewmanBuildingMicroservices}.
After gaining essential insights of \focus\textit{'} domain, we employed a static software structure analysis tool to map source code packages to the previously identified target boundaries.
This enabled us to find overlaps among business functions for identifying ambiguities in the corresponding service boundaries~\cite{evans2004ddd,FowlerUbiquitousLanguage}.
We then used dynamic analysis and live visualizations of \focus\textit{'} runtime behavior to refine previously identified service boundaries and find actual microservice candidates.

The remainder of this paper is structured as follows.
Section~\ref{sec:related-work} presents related work.
Section~\ref{sec:ddd} introduces the domain analysis of \focus.
In Section~\ref{sec:static-analysis}, we build upon gathered domain knowledge and statically analyze \focus\textit{'} source code.
Section~\ref{sec:dynamic-analysis} presents the procedure and results of the dynamic analysis of \focus\textit{'} runtime behavior.
Finally, we discuss our main conclusions in Section~\ref{sec:conclusions}.

\section{Related Work}\label{sec:related-work}
There exist several approaches, which are related to our migration process, based on the methodology or research topic.  

Gysel et al.~\cite{Gysel2016} utilize the static software system and domain analysis tool \textit{Service Cutter}\footnote{\url{https://servicecutter.github.io}} to discover a suitable decomposition into microservices within the software architecture of the cargo tracking application \textit{Cargo Tracker}.\footnote{\url{https://cargo-tracker.gitbook.io}}
The decomposition process is based on several coupling criteria from literature and industrial experience.
They extract coupling information from software artifacts of the software system like use cases and create an undirected, weighted graph to identify clusters.
These clusters serve as candidates for the decomposition.
Baresi et al.~\cite{Baresi2017} also performed a tool-supported domain analysis on \textit{Cargo Tracker}.
In contrast to Gysel et al.\, they aim to provide an automated solution for decomposing a given domain into candidate microservices with the help of a pre-computed database of collections and similar words.
The goal is to transform the domain into a machine-readable format to find cohesive groups.
In comparison to both approaches, we focus on a collaborative approach with the software developers to create a context map, identify related BCs, and finally present potential candidate microservices.
Additionally, these microservice candidates are verified and further decomposed due to applied static and dynamic analyses.

In~\cite{KnocheMicroservices:2018}, the authors present a process for decomposing an existing software asset into microservices.
The process is based on experience from an industrial case study involving the migration of a COBOL legacy system towards Java.
Their proposed modernization process consists of five steps, starting with defining external service facades and finally replacing service implementations with microservices.
In contrast to our approach, we create a context map instead of a domain model and discuss use cases of the software system with the developers to identify BCs and microservice candidates.

A vertical decomposition of a monolith into several, self-contained systems is presented in~\cite{OttoMicroservices:2017}.
The authors report on their successful decomposition and appropriate granularity of microservices as well as coupling, integration, scalability, and monitoring of microservices at the e-commerce platform \textit{otto.de}.
Compared to our approach, they focus on the technical and organizational aspects of the migration process and thus also address the development and deployment process.
We employ a DDD-based approach for a microservice decomposition and do not (yet) focus on quality aspects of the migration. 

In~\cite{Fan2017}, the authors proposed a migration process for moving from a monolithic to a microservice architecture and applied it on a mobile learning application.
Similar to our approach, they extracted microservice candidates from the original system based on DDD.
In the next step, they determine, whether the database schema within the software system is consistent with the microservice candidates and the exclusion of inappropriate candidates.
Afterwards, they extract source code related to the microservice candidates and establish a communication between them.
Finally the applied decomposition is tested and deployed in the execution environments.
Compared to our approach, we additionally apply static and dynamic analysis to gather supplemental, valuable information of identified bounded contexts to optimize and further decompose our potential candidate microservices.

Our employed live trace visualization tool \explorviz\ uses a 3D city metaphor visualisation for software applications. Related approaches employ this metaphor, for instance, for 
analyzing memory leaks~\cite{Weninger2019} or synchronization problems~\cite{SynchroVis2013}. Different to these approaches, we use the city metaphor for refining and revising the microservice decomposition obtained from static analysis. 
\section{Domain-driven design}\label{sec:ddd}
\begin{figure}[htbp]
	\centerline{\includegraphics[width=0.65\linewidth]{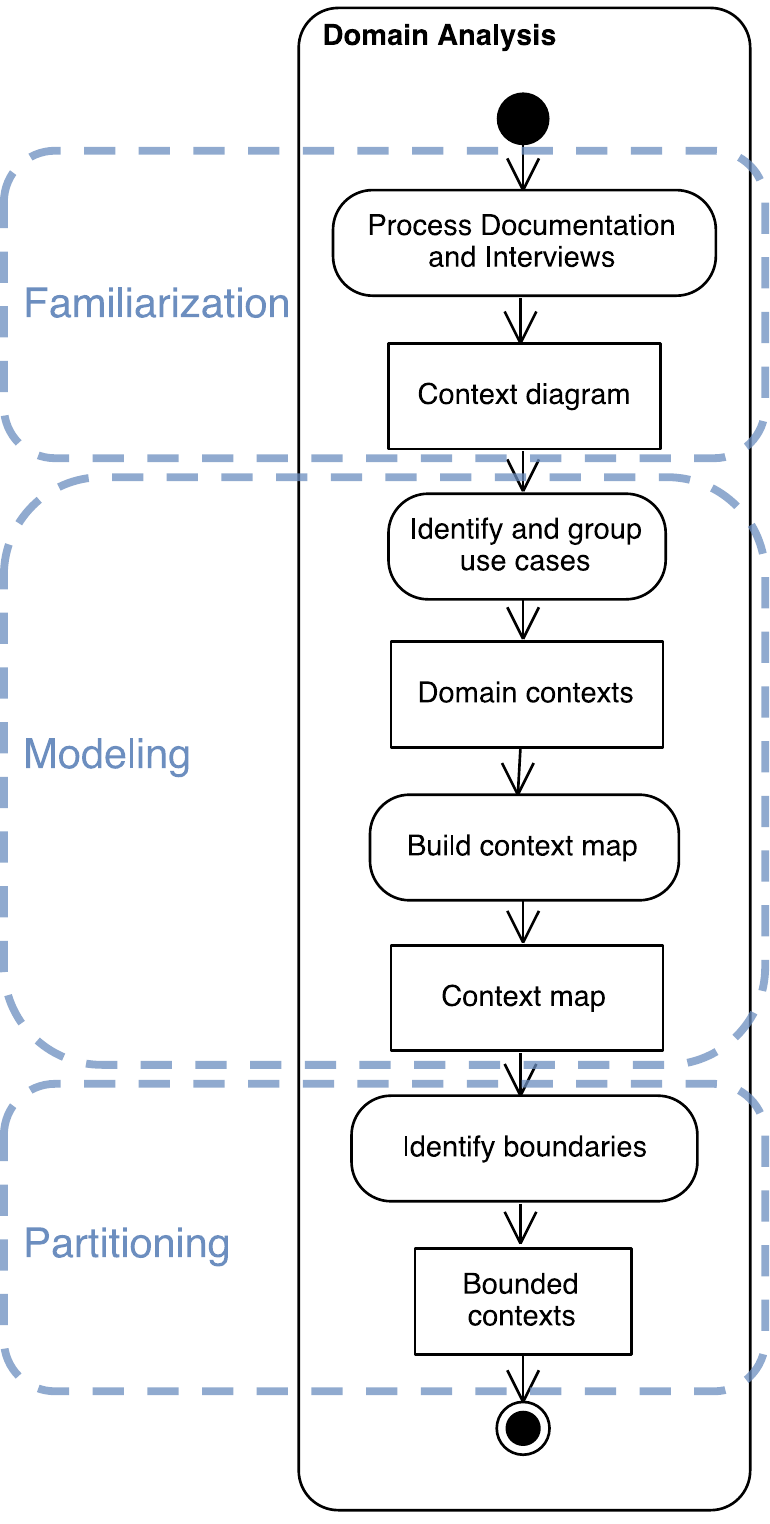}}
	\caption{Domain analysis procedure with actions (rounded rectangles), intermediate and resulting data (rectangles), and phases (blue boxes).}
	\label{fig:domain-analysis-procedure}
\end{figure}

Fig.~\ref{fig:domain-analysis-procedure} depicts all actions, as well as intermediate and resulting artifacts of our analysis modeled as a Unified Modeling Language (UML) activity diagram.
The domain analysis of \focus\ was divided in three phases, as indicated with the dashed boxes.

The \textit{Familiarization} phase was used to gather information about \focus\ and to recover terms of its ubiquitous language.
This was a crucial step, since we did not have any knowledge about the software, its structure, or behavior.
\newpage \noindent For that reason, \adesso\ supplied us with documentation, e.g., customer requirements specifications.
Furthermore, we were able to interview developers and domain experts of \focus.
A first result of this phase was a context diagram for the domain actors, i.e., \textit{lottery application}, \textit{user}, \textit{customer}, and \textit{state lottery}.
\textit{Customers} of \focus\ represent a subset of \textit{users} who actively engage with the offered products.
Remaining actors were summarized in a group called \textit{other}, e.g., the OASIS system which logs every banned player to counteract gambling addiction.
We used the actor context diagram in additional interviews to discover more of \focus\textit{'} internals and behavior.
The collected information of the \textit{Familiarization} phase provided us with domain knowledge, which served as foundation for the next activities in the analysis.

The \textit{Modeling} phase was used to clarify the behavior of \focus\ and derive an architectural overview of the system.
With the help of developers and domain experts, we first identified the use cases of the \textit{lottery application} actor towards the remaining actors.
Then, we defined domain contexts, i.e., groups for related behavior or structure, and mapped the use cases onto these.
We observed that some use cases related to multiple domain contexts.
For example, the use case \textit{transfer money to online wallet} was mapped both onto the \textit{Payment Method} and \textit{Online Wallet} contexts.
This indicated a potential ambiguity, which was later reviewed in the static analysis (see Section~\ref{sec:static-analysis}).

\begin{figure}[htbp]
	\centerline{\includegraphics[width=\linewidth]{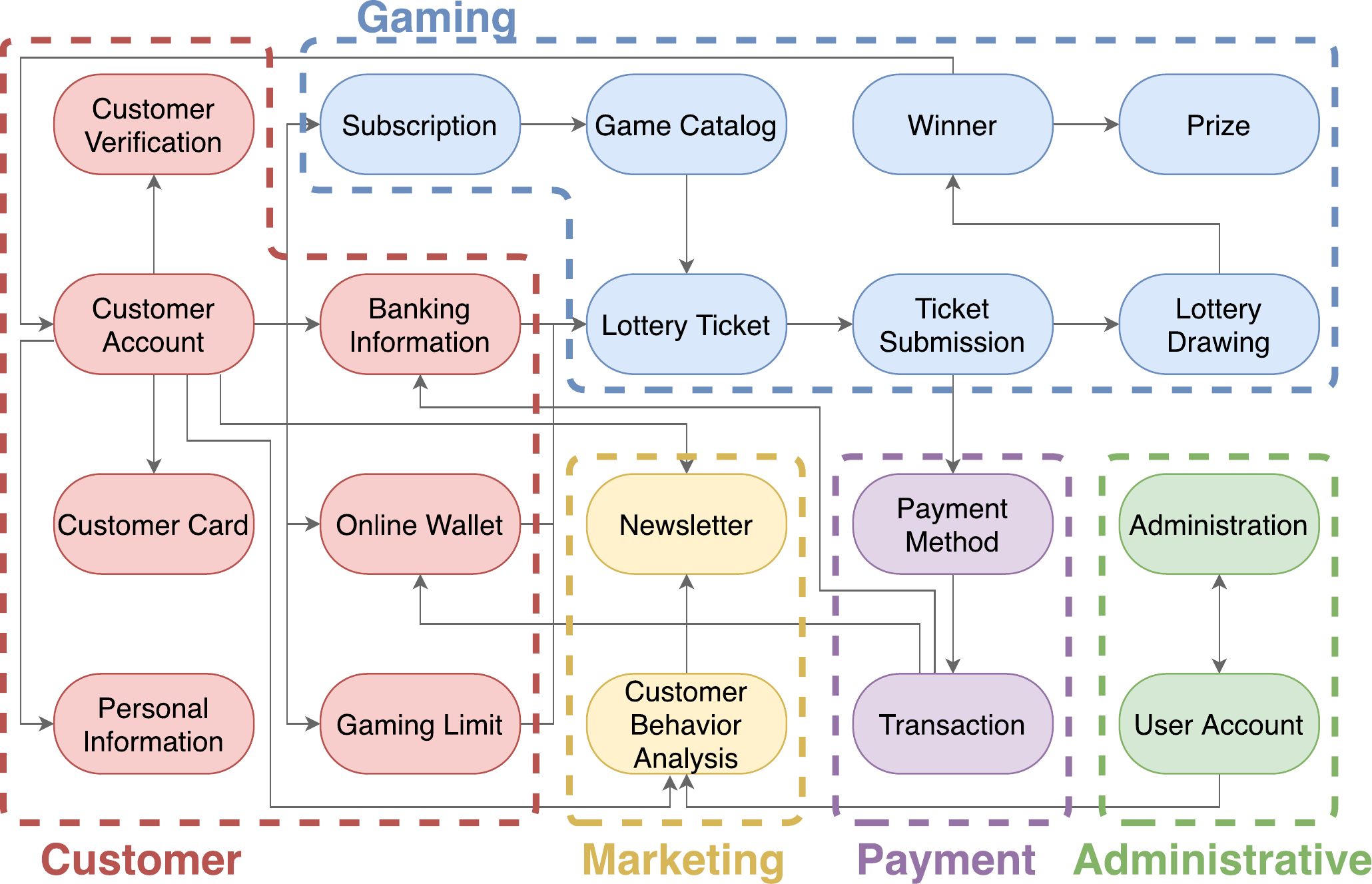}}
	\caption{Partitioning of the context map based on suitable boundaries (in color and dashed boxes) results in bounded contexts. A single rounded rectangle represents grouped use cases, i.e., a domain context.}
	\label{fig:partitioning-context-map}
\end{figure}

The \textit{Familiarization} phase revealed relationships between actors.
Since the actors were related to the use cases and these were mapped to domain contexts, we proceeded by denoting the relationships at domain context level.
The resulting context map with the defined domain context and their relationships can be seen in Fig.~\ref{fig:partitioning-context-map}.

In the final \textit{Partitioning} phase, we used the collected domain knowledge to identify target boundaries in the context map.
These boundaries represent scopes in which each domain term has a unique definition, i.e., bounded contexts.
Bounded contexts are the foundation for the microservice decomposition process~\cite{NewmanBuildingMicroservices,BoundedContextVsMicroservices}.
Equipped with the multiple domain context conflicts of the use cases in the \textit{Modeling} phase, we were able to validate our chosen boundaries and resolve potential ambiguity at implementation level via static analysis (see Section~\ref{sec:static-analysis}).

\section{Static Analysis}\label{sec:static-analysis}
\begin{figure}[tbp]
	\centerline{\includegraphics[width=\linewidth]{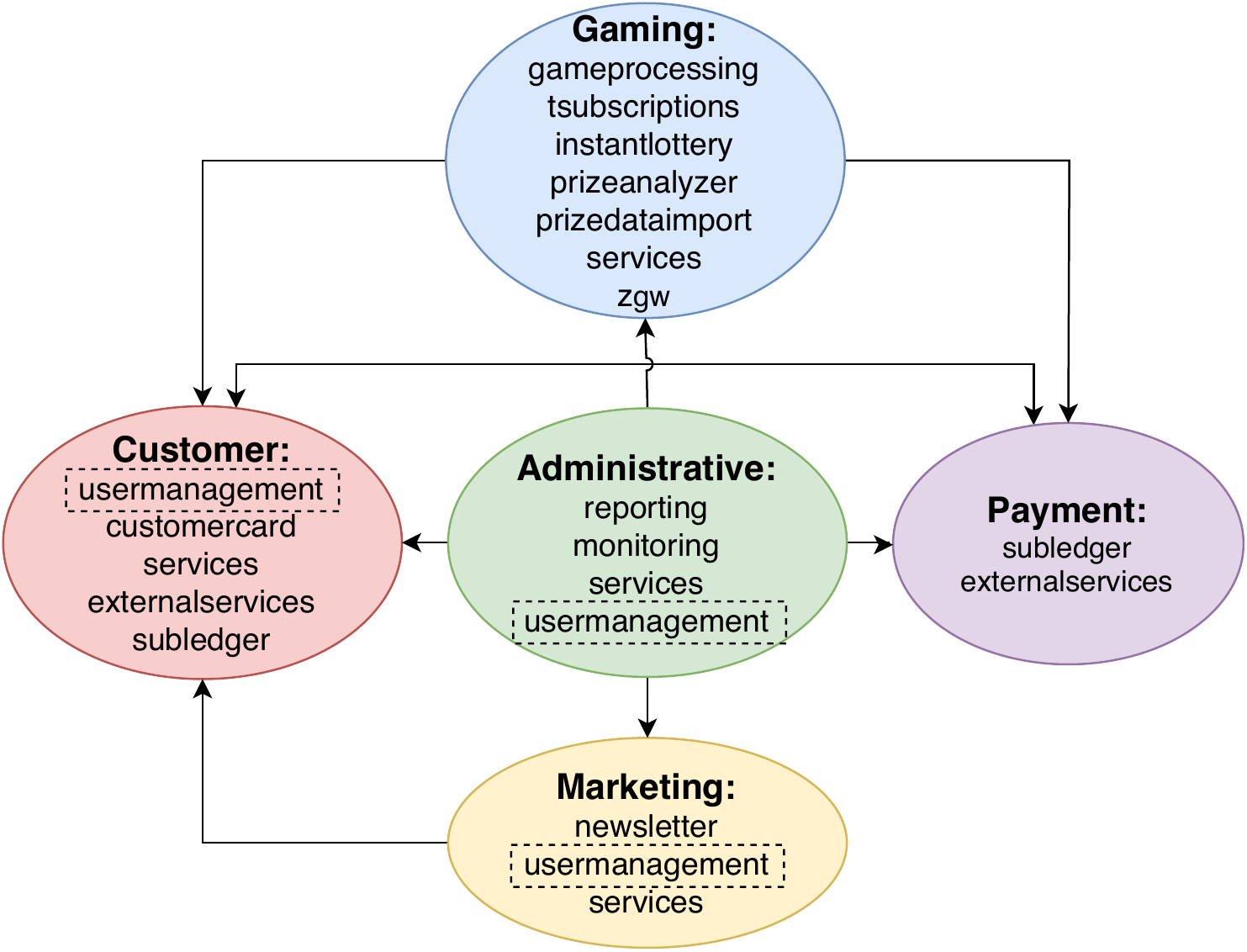}}
	\caption{As a result of the static analysis, source code packages were mapped to bounded contexts (bold). Dashed boxes indicate remaining ambiguities at implementation level.}
	\label{fig:bc-source-mapping}
\end{figure}

After decomposing \focus\ at the conceptual domain level, the next step was to partition its source code.
For that, we employed the static software structure analysis tool \textit{Structure101}\footnote{\url{https://structure101.com/}} to analyze the source code packages (SCP).
\textit{Structure101's} levelized structure maps enabled us to observe dependencies among the SCPs and identify thier locations at implementation level.
We manually mapped the SCPs to the identified use cases of the \textit{Modeling} phase (see Sec.~\ref{sec:ddd}).
Since we already assigned the use cases to the domain contexts (see Sec.~\ref{sec:ddd}) as well as partitioned them into bounded contexts (Fig.~\ref{fig:partitioning-context-map}), we obtained a direct mapping of the \focus\ SCPs to the selected bounded contexts.
This led to a first decomposition of the \focus\ application.
Fig.~\ref{fig:bc-source-mapping} illustrates the assignment of the SCPs to the bounded contexts.

The \textit{Customer} context contains the main functionality of the customer account management that is provided by the package \texttt{usermanagement}. 
This includes the account creation, login and logout, as well as editing personal information.
For these use cases, identity verification services and external OASIS checkups are made accessible by the SCPs \texttt{services} and \texttt{externalservices}.
A customer card contains the required customer account information for offline gaming at local lottery offices. 
The games played with the customer card are then booked on the online customer account.
However, a customer card can exist without ever being used for gaming and can be seen as part of the customer account.
Thus, we assigned the customer card management to the \textit{Customer} context, rather than to the \textit{Gaming} context.
Since ordering and paying for a new customer card is also part of the customer card management, certain functions of the \texttt{subledger} package are also part of the \textit{Customer} context.
To complete the payment process, the \textit{Customer} context is dependent on the \textit{Payment} context.

The \textit{Gaming} context covers the gaming functionality of \focus. 
The customer chooses a lottery game from an offered game catalog, fills out a lottery ticket and creates a game order (\texttt{gameprocessing}). 
The assignment to the customer account creates the indicated dependency between \textit{Gaming} and \textit{Customer}.
\focus\ also provides instant lotteries such as online scratch tickets (\texttt{instantlottery}). 
These instant lotteries are immediately drawn. 
The conventional lottery games however are drawn at a certain point in time. 
Thereafter, the drawing results are imported (\texttt{prizedataimport}) and the individual lottery prize of each winner is calculated (\texttt{prizeanalyzer}).
Alternatively, the customer can also subscribe to specific games (\texttt{tsubscriptions}). 
A subscribed game is repeatedly played for a defined period of time.
\newpage \noindent Of course, the lottery tickets and the subscriptions have to be paid for. 
Therefore, the gaming process is dependent on the \textit{Payment} context. 
The same dependency can be found in the package \texttt{zgw}, which is a German acronym for central winnings management. 
It manages the customer winnings. 
Winnings are either transferred back to the personal online wallet or to the bank account of the customer.

The \textit{Payment} context handles and takes account of all incoming and outgoing money transactions (\texttt{subledger}).
Additionally, it provides various payment methods to the customer (\texttt{externalservices}).
To receive the up-to-date banking information or the online wallet ID of the customer, \textit{Payment} is dependent on \textit{Customer}.

The \textit{Marketing} context provides newsletters to which the customers can subscribe. 
As long as the customer is subscribed to a newsletter, a notice, either through the website or via e-mail (\texttt{services}), is delivered. 
Lottery operators can create or edit newsletters. 
To promote these newsletters to potentially interested (new) customers, they can define a target group to which the newsletter is presented more vividly. 
This promotion process creates the dependency between \textit{Marketing} and \textit{Customer}.

The \textit{Adminstrative} context comprises the reporting and monitoring services of \focus. 
The customer activity as well as the performance and the load of the software system are monitored. 
System reports are created continuously and distributed to the appropriate authority (\texttt{services}). 
Furthermore, this context manages the user accounts and their rights (\texttt{usermanagement}). 
The monitoring of the system’s activity creates the dependencies to all other contexts of \focus.

After mapping the SCPs to the bounded contexts, we were able to validate our chosen boundaries and to identify potential ambiguities at implementation level.
The domain context conflicts of the use cases in the \textit{Modeling} phase (see Sec.~\ref{sec:ddd}) already indicated required restructurings of some boundaries.
One example for this can be seen in Fig.~\ref{fig:bc-source-mapping}:
\texttt{usermanagement} was mapped to multiple bounded contexts due to shared code.
This disclosed a required restructuring of this package to achieve well-defined service boundaries for a good microservice decomposition~\cite{NewmanBuildingMicroservices}.

\begin{figure*}[btp]
	\centering 
	\subfigure[Asynchronous communication.]{\label{fig:a}\includegraphics[width=45mm]{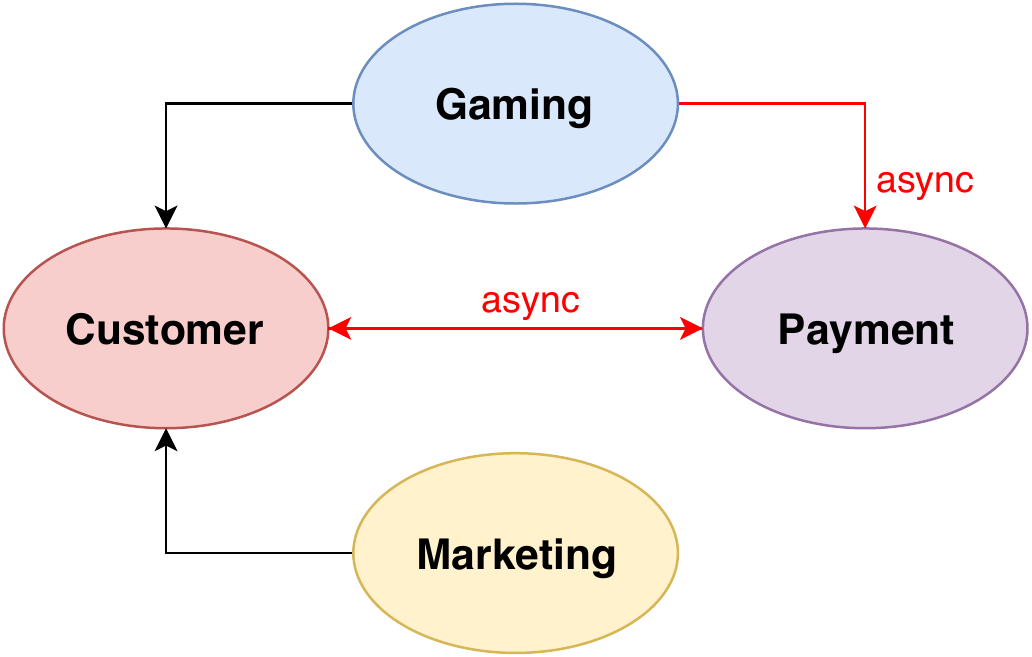}}
	\subfigure[Merging.]{\label{fig:b}\includegraphics[width=43mm]{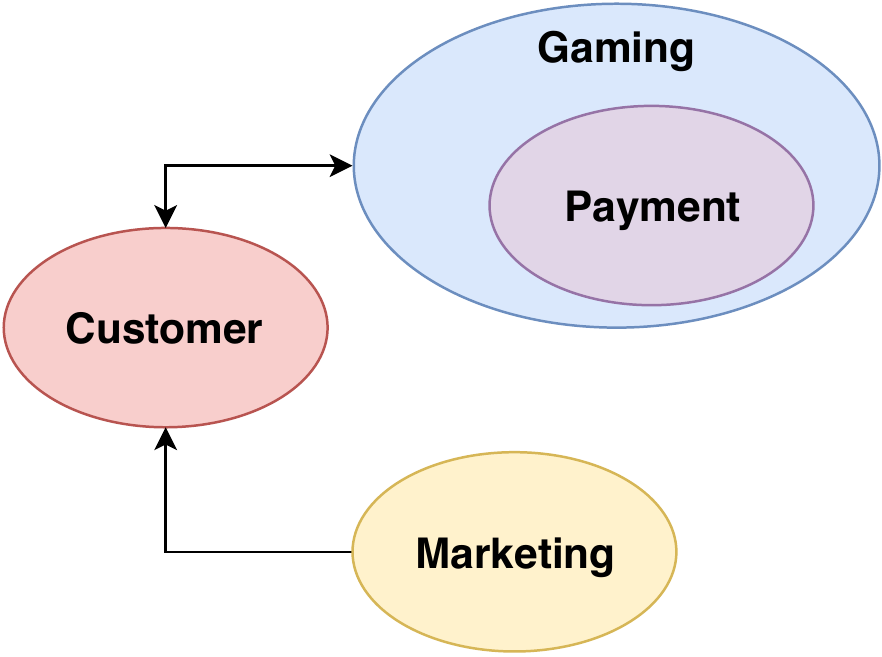}}
	\subfigure[Dividing \textit{Gaming}.]{\label{fig:c}\includegraphics[width=45mm]{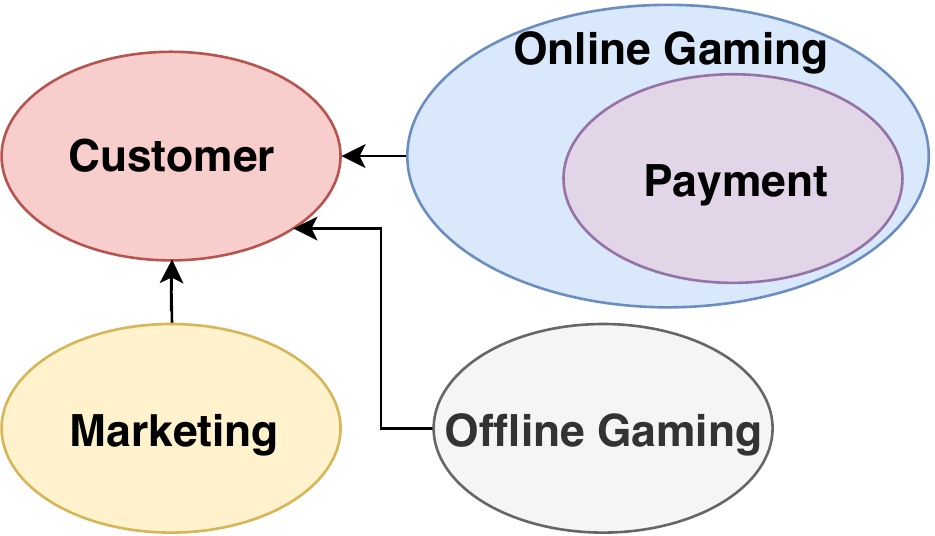}}
	\subfigure[Multiple \textit{Payment} components.]{\label{fig:d}\includegraphics[width=45mm]{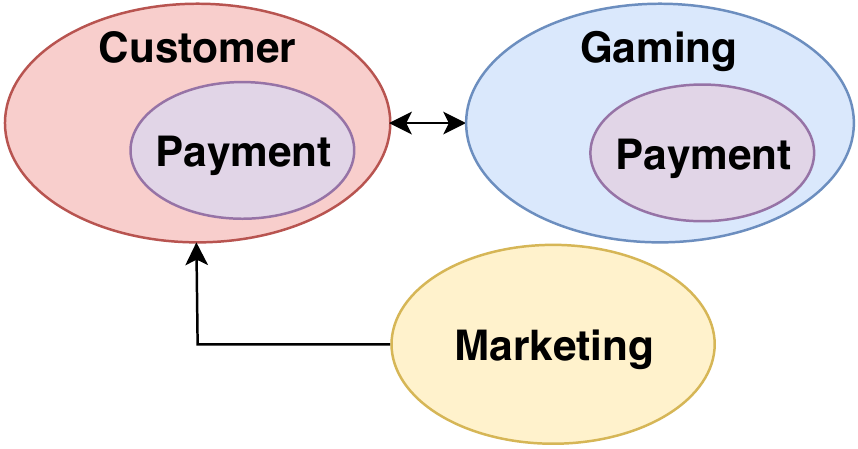}}
	\caption{Four architectural alternatives for addressing the dependency problem with the \textit{Payment} context from Fig.~\ref{fig:bc-source-mapping}.}
	\label{fig:payment-solutions}
\end{figure*}

Fig.~\ref{fig:bc-source-mapping} revealed another problem:
both, the \textit{Customer} and the \textit{Gaming} contexts were dependent on the \textit{Payment} context.
This could become a problem, if we intended to derive microservices from the bounded context-based decomposition.
For example, a customer could not successfully acquire a customer card or play a lottery game when the microservice for the \textit{Payment} context (also called \textit{Payment service}) is unavailable.
One could argue that buying a customer card is not one of the primary functionalities and that a temporary system failure is tolerable.
Therefore, it would be acceptable to asynchronously forward the buying request to the \textit{Payment} service.
However, this approach would be inappropriate when it comes to playing lottery games.
In this case, it would be preferable to successfully play a game without any dependencies to other services. 

Fig.~\ref{fig:payment-solutions} presents four identified architectural alternatives to address this problem.
Figure~\ref{fig:a} depicts the asynchronous communication solution.
The rationale for this approach is the redefinition of success for the use case of filling out and submitting a lottery ticket in the \textit{Gaming} context. 
Beforehand, it was considered to be completed after the game order of the lottery ticket was created and paid for. 
As an alternative, we can consider the use case for the \textit{Gaming} context to be completed right after the game order has been created. 
Completing the payment process is then part of the \textit{Payment} context. 
The game order fails when the payment process has not been completed until the ticket submission deadline of the lottery game. 
In this case, the \textit{Payment} service informs the \textit{Gaming} service. 
Therefore, an asynchronous communication protocol between \textit{Payment} and the other contexts or services would suffice. 
As a result, the dependencies between the services do not change.
However, each service can now independently handle its own use cases.

Due to space limits for this paper, we only explain the further investigation of the finally chosen solution (a) in detail in the following Section~\ref{sec:dynamic-analysis}.
The other solutions were eventually declined due to service granularity (b and c) and synchronization (d) problems.

\section{Dynamic Analysis}\label{sec:dynamic-analysis}
We expanded on the results of the static analysis with the help of our trace visualization tool \explorviz.\footnote{\url{https://www.explorviz.net/}}
\explorviz\ enables a live monitoring and visualization of large software landscapes~\cite{ExplorViz}.
In particular, the tool offers two types of visualizations -- a landscape-level~\cite{ECIS2015} and a 3D application-level perspective.
The first provides an overview of a monitored software landscape consisting of several servers, applications, and communication in-between.
The second perspective visualizes a single application within the software landscape and reveals it's underlying architecture, e.g., the package hierarchy in Java, and shows classes and related communication. 
The tool has the objective to aid the process of system and program comprehension for developers and operators.
It has been empirically evaluated via controlled experiments~\cite{VISSOFT2015hierarchical,ExplorVizControlledExperiment2015}. \explorviz\ started in 2012 as a layered, monolithic web application. Meanwhile, we modularized ExplorViz itself into a microservice architecture for improved collaborative open-source development~\cite{COLLA2019}.
In the present paper, we employ the 3D application-level perspective of \explorviz\ for modularizing \focus\ into a microservice architecture.

By dynamically analyzing the behavior of \focus\ with \explorviz, we refined the bounded contexts.
This analysis of \focus\ has the following two goals: First, we further investigated the presented asynchronous solution for the \textit{Payment} problem (see Figure~\ref{fig:payment-solutions}).
Second, we intended to discover additional microservice candidates within the software architecture to further decompose the resulting software architecture (see below).

\begin{figure*}[tb]
	\centering%
	\includegraphics[width=\textwidth]{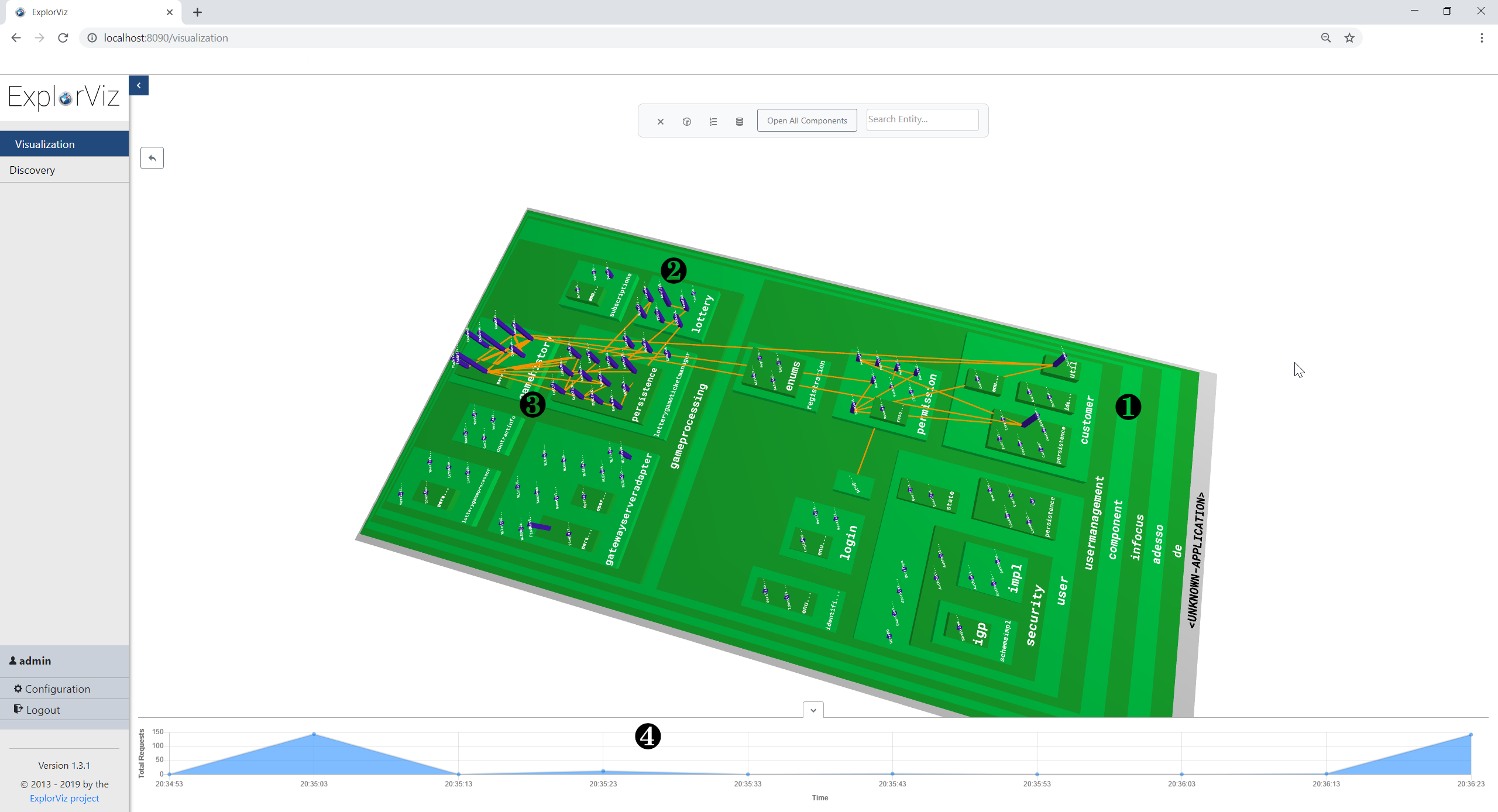}%
	\caption{Overview snapshot of an excerpt of \focus\textit{'}~runtime behavior with the \explorviz\ 3D visualization.}
	\label{fig:explorviz-infocus}%
\end{figure*}%

Figure~\ref{fig:explorviz-infocus} presents an overview snapshot of \focus\textit{'} runtime behavior visualized with \explorviz.
\explorviz\textit{'} 3D visualization is based on the city metaphor.
SCPs are depicted as green boxes (\one).
Classes are visualized as purple bars (\two), whereby the height of a class is related to the instance count at runtime.
Communication between these entities is shown by orange lines (\three).
The width of a communication line is related to the number of requests in that visualized snapshot.
\explorviz\ provides a timeline (\four), such that users can go back to previous visualized runtime behavior snapshots.

\begin{figure*}[tb]
	\centering%
	\includegraphics[width=\textwidth]{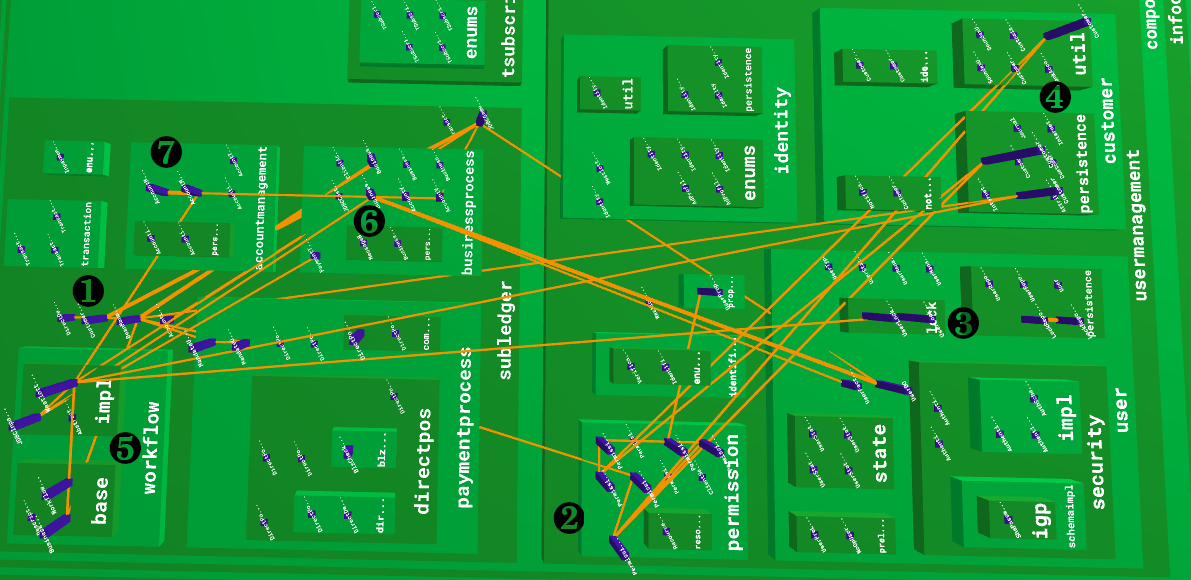}%
	\caption{Monitoring and visualizing \focus\textit{'} runtime behavior with \explorviz\ when a customer transfers money to his online wallet.}
	\label{fig:explorviz-transer-money}%
\end{figure*}

We continued the investigation of the asynchronous solution to the \textit{Payment} problem as shown in Figure~\ref{fig:a}.
The goal was to achieve decoupling via asynchronous communication between the \textit{Customer} and \textit{Gaming} services and the \textit{Payment} service.
Since the current definition of the service boundaries via bounded contexts lead to tight coupling of the three mentioned services when it comes to buying a product of the lottery application, we needed to redraw these boundaries.
Figure~\ref{fig:explorviz-transer-money} shows the runtime behavior when a customer transfers money from his registered bank account to his online wallet.
This use case is related to all three services.
The execution can be broken down into seven distinct steps:\\

\paragraph*{Step 1} These classes (\one in Figure~\ref{fig:explorviz-transer-money}) are the centerpiece of this use case and manage the payment process.

\paragraph*{Step 2} The required permissions for transferring money to the target online wallet are checked (\two in Figure~\ref{fig:explorviz-transer-money}).

\paragraph*{Step 3} Checks of user rights and status are conducted in the \texttt{user} package (\three in Figure~\ref{fig:explorviz-transer-money}).

\paragraph*{Step 4} The \texttt{customer} package (\four in Figure~\ref{fig:explorviz-transer-money}) delivers the required information on the customer and the associated online wallet.

\paragraph*{Step 5} Inside the \texttt{workflow} package (\five in Figure~\ref{fig:explorviz-transer-money}), the processing of the selected payment method is managed.

\paragraph*{Step 6} The \texttt{businessprocess} package (\six in Figure~\ref{fig:explorviz-transer-money}) is responsible for managing the appropriate business process. 
For that, business records are created to track the respective process.

\paragraph*{Step 7} The account of the customer online wallet, onto which the money is transferred to, is managed (\seven in Figure~\ref{fig:explorviz-transer-money}).
\\\\
To reduce the dependencies between \textit{Customer} and \textit{Payment} while enabling a feasible asynchronous communication between these services, we redefine the \textit{Payment} service into an \textit{Order} service.
This service handles the payment process as well as the shopping cart which contains the items to buy.
The shopping cart was previously part of the \textit{Customer} service.
This previous assignment of the shopping cart, however, led to unwanted communication between the \textit{Gaming} service when buying lottery tickets.
With the new assignment of the shopping cart component, both the \textit{Customer} and \textit{Gaming} services signal the \textit{Order} service asynchronously to put a requested product into the cart.
This can be done through choreographed publish and subscribe events.
This approach has the advantage that all products can still be viewed or requested to buy even when the \textit{Order} service is not available at that time. 
Furthermore, no more inter-service communication has to take place to complete the payment process. 
The other services can then be informed via asynchronous communication to confirm the successful execution of the payment process.

\begin{figure}[htbp]
	\centerline{\includegraphics[width=\linewidth]{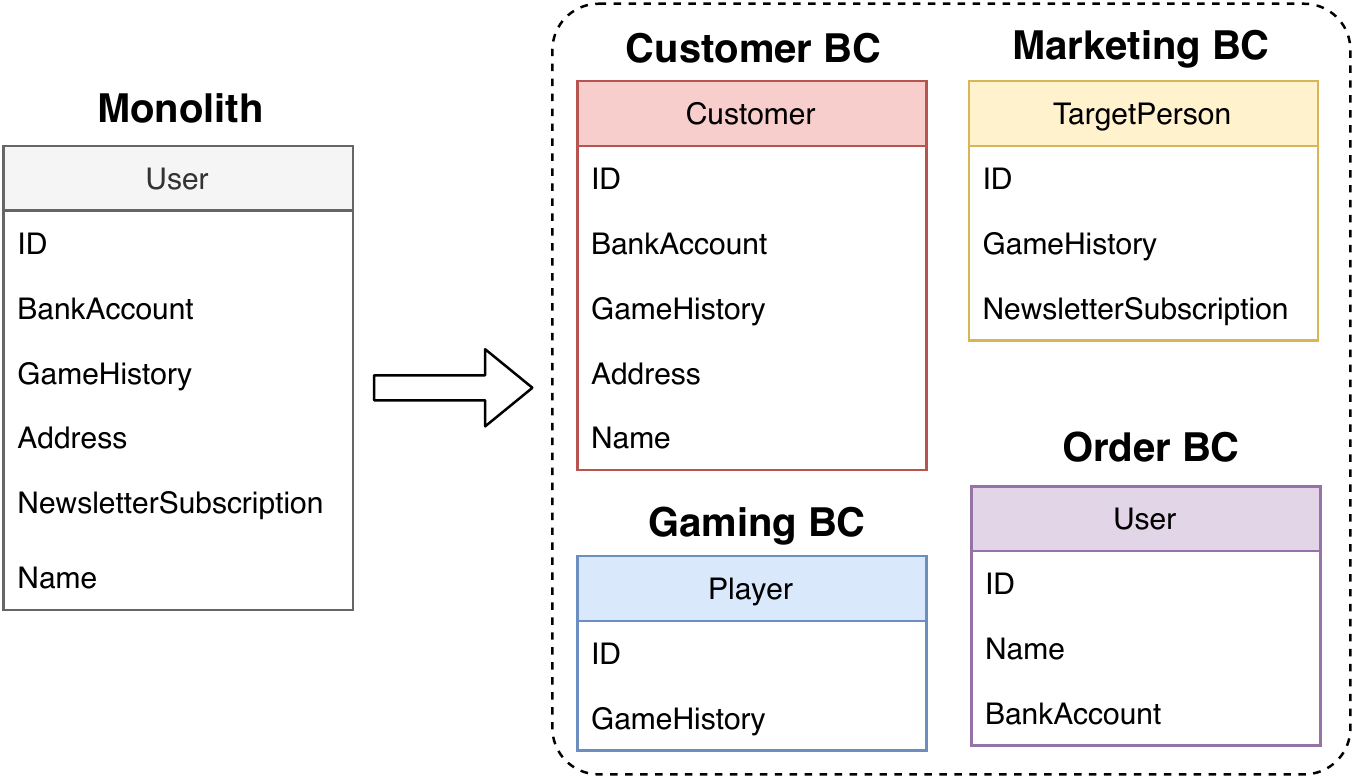}}
	\caption{Partition of the monolithic data model \textit{User} into separate data models for each bounded context. The \textit{Payment} bounded context is replaced by the \textit{Order} bounded context.}
	\label{fig:data-division}
\end{figure}

An important property of microservice architectures is that each microservice manages its own data store, possibly with different database technologies (polyglott persistence~\cite{NoSQL2012}).
Consequently, the next step of our analysis process was to investigate a possible partition of \focus\textit{'} data model.
Therefore, the centralized data governance had to be replaced by a distributed data model.
For \focus, we statically analyzed the data model alongside previously identified use cases.
We employed the database administration tool \textit{DBeaver}\footnote{\url{https://dbeaver.io/}} to identify the database tables which are used by the use cases.
Then, we mapped the tables onto the respective bounded contexts.
This investigation of transaction boundaries and the resulting assignment provided us with an individual schema for each service~\cite{NewmanBuildingMicroservices}.
Afterwards, the tables were adjusted to their context.
Some tables, however, were used by multiple services.
One example for that can be seen in Figure~\ref{fig:data-division}.
The previous \textit{User} table of the legacy system was used by multiple business functions, even though the definition of a user depends on its business context.
For example, in contrast to the \textit{Customer} service that considers users as customers, the \textit{Gaming} service considers users as players of a game. These different views are now reflected in the separate data models for each bounded context in Figure~\ref{fig:data-division}.\\

\noindent To discover additional microservices, we executed the identified use cases and analyzed the resulting 3D visualization.
Figure~\ref{fig:explorviz-customer-data} shows an area of the runtime behavior of \focus\ when a lottery ticket is selected by a user and completed by the application.
This use case can be divided into four distinct steps:

\begin{figure}[htbp]
	\centerline{\includegraphics[width=\linewidth]{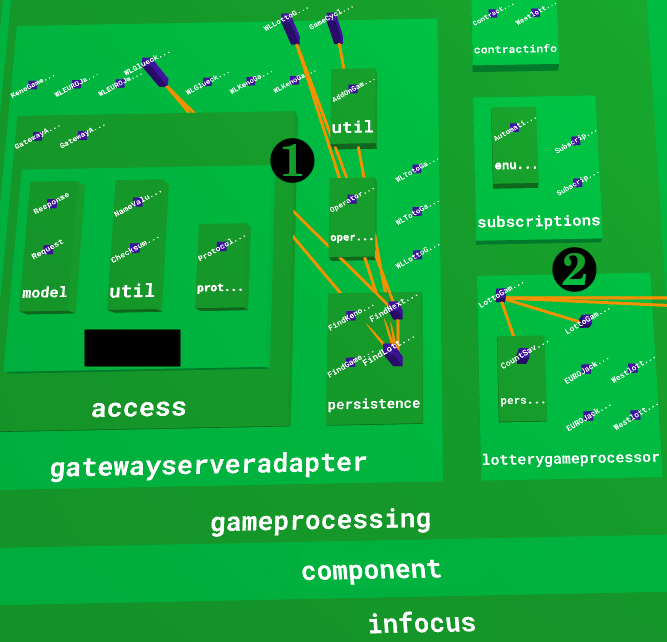}}
	\caption{Visualization of \focus\textit{'}~runtime behavior with \explorviz\ when loading and filling out a lottery
ticket as user.}
	\label{fig:explorviz-customer-data}
\end{figure}

\paragraph*{Step 1} Inside the \texttt{gatewayserveradapter} package (\one in Figure~\ref{fig:explorviz-customer-data}), a lottery manager is invoked for the appropriate lottery game which was selected by the customer.
Required lottery information, such as drawing dates, are collected and the lottery ticket is created.
Furthermore, the delivery of the required lottery ticket information is done by the internal functionality of the persistence sub-package to assemble the actual ticket.
These attributes can represent different available jackpots or gaming options for the selected game.

\paragraph*{Step 2} The classes of the \texttt{lotterygameprocessor} package (\two in Figure~\ref{fig:explorviz-customer-data}) manage the gaming process of the selected lottery game.

\paragraph*{Step 3} (not shown) Customer data, such as the ID and the personal gaming limits, are obtained. 
These are required for successfully playing a lottery game as customer.

\paragraph*{Step 4}  (not shown) The \texttt{lottery} package manages the lottery ticket. It tracks the selected fields and specific gaming strategies of the played lottery ticket.
\\\\
We can see that the use case can be split into two main phases.
First, the selected lottery ticket is assembled by collecting the gaming information. 
Thereafter, the customer fills out the ticket and selects different options for his game. 
Here, an opportunity can be found to define a microservice. 
This microservice, which we call \textit{TicketManager}, handles the management and the composition of lottery game attributes for the creation of a specific lottery ticket.
Hence, it holds all the available options for each lottery game.
Depending on the customer's choice, it assembles the necessary information for creating a lottery ticket that the customer can then fill out.
Therefore, the \textit{TicketManager} microservice encapsulates the functionality of the first step (\one in Figure~\ref{fig:explorviz-customer-data}).
The introduction of this microservice enables the developers to easily define and add new games to the application.
Furthermore, this part of the gaming process can be seen as a potential bottleneck, depending on the simultaneously requested games.
Therefore, the now achieved option of horizontal scaling is desirable.
The \textit{TicketManager} service additionally needs to submit the assembled attributes to the \textit{Gaming} service which handles the further gaming process.
One possibility is the asynchronous communication between these services by publishing an6d subscribing to events.

\section{Conclusions}\label{sec:conclusions}
In this paper, we present experience with our migration approach that extends static analysis with dynamic analysis of a software system's runtime behavior to support the decomposition of a legacy, monolithic architecture into microservices.
We combined established analysis techniques for microservice decomposition, such as the bounded context pattern of domain-driven design, and refined the architectural design via dynamic software visualization to identify appropriate microservice candidates.
Especially employing runtime software visualization using the live trace visualization tool \explorviz\ proved to be a valuable step within our migration approach.
Thus, we were able to confirm and refine the service boundaries which were obtained from the initial domain analysis.

We successfully applied our approach in collaboration with the German IT service provider \adesso\ to their real-word, legacy lottery application \focus\ to identify potential microservice decompositions for their existing layered monolithic Enterprise Java system.
The redefinition of the \textit{Payment} service and the discovery of the additional \textit{TicketManager} service showcase the benefits of dynamic analysis for microservice decomposition in Section~\ref{sec:dynamic-analysis}.
By monitoring the runtime behavior of \focus, we gained new insights which were previously overlooked by the static analysis (see Section~\ref{sec:static-analysis}).
The development team of \focus\ will use our results as foundation for a future decomposition into microservices.

After applying our approach to a real-world application for the first time, we now can derive certain lessons learned for other practitioners and researchers.
Of course, the domain analysis is key for the overall approach.
We incrementally refined intermediate and resulting data of the analysis with the help of documentation and discussions.
Furthermore, we internally explained the domain to each other.
This highlighted uncertainty that was then reinvestigated.

In the future, we plan to improve our approach and apply it to aid the migration process of additional software systems in collaboration with industrial partners.\footnote{During the migration process, we employed the tools \textit{Structure101}, \textit{ExplorViz}, and \textit{DBeaver}. These tools are publicly available on the related websites. Corresponding documentation may help other researchers or practitioners to replicate our process in a similar setting.}
Specifically, we will investigate how we can consider design issues for microservices, e.g., code replication and non-functional requirements such as team size or security.
Additionally, we enhance \explorviz\ capabilities regarding filtering executed traces to better analyze specific sequences of communication within the observed software system.
Moreover, we will extend its 3D software visualization towards collaborative Virtual Reality to offer an immersive user experience and natural interaction based on~\cite{VISSOFT2015VR,ExplorVizVR}.

\printbibliography

\end{document}